# Buried Interfaces and Spin Orientation in [Co/Pt]$_{10}$/Fe multilayer with Orthogonal Magnetic Anisotropy: Effect of Fe Thickness

Sadhana Singh[1], Manisha Priyadarsini[1], Sharanjeet Singh[1], Ilya Sergeev[2], Marcus Herlitschke[2], H. C. Wille[2], Dileep Kumar[1*]

[1]UGC-DAE Consortium for Scientific Research, Khandwa Road, Indore-452017, India
[2]Deutsches Elektronen-Synchrotron DESY, Notkestraße 85, 22607 Hamburg, Germany
[*]Corresponding Author Email: dkumar@csr.res.in; dileep.esrf@gmail.com

**Abstract.** In the present work, spin orientation and variation of the strength of coupling in [Co/Pt]$_{ML}$/Fe multilayer have been investigated as a function of the thickness of the Fe layer. [Co/Pt]$_{ML}$/Fe multilayer has orthogonal anisotropy with the Fe layer, and [Co/Pt]$_{ML}$ has in-plane magnetic anisotropy and perpendicular anisotropy, respectively. Measurements are performed using in-situ magneto-optical Kerr effect (MOKE) and isotope-sensitive depth-resolved nuclear resonance scattering (NRS) technique. Real-time in-situ MOKE measurement during Fe growth reveals that with an increase in thickness of the Fe layer, moments of Fe layer reorientation from out of the plane to in-plane direction. This is attributed to the decrease in the coupling between the [Co/Pt]$_{ML}$ and Fe layer. For the depth-dependent study, two [Co/Pt]$_{ML}$/Fe multilayers having the same thickness but different positions of the Fe$^{57}$ marker layer ([Co/Pt]$_{ML}$Fe/Fe$^{57}$) and [Co/Pt]$_{ML}$Fe$^{57}$/Fe ) were studied using the NRS technique. Films with varying external magnetic fields were also studied to investigate coupling strength. Measurements were performed under the x-ray standing wave conditions to enhance resonance yield. It is observed that for the 75 Å Fe in [Co/Pt]$_{ML}$/Fe multilayer, the coupling varies along the depth of the Fe layer. The coupling is strong at [Co/Pt]$_{ML}$ and Fe interface with spins of the Fe layer aligned in the out-of-plane direction, whereas moments away from the interface are weakly coupled and aligned in-plane along the magnetic easy axis. Due to this gradient in strength of coupling along the depth, a large magnetic field is required to reorient spins at the interface along the magnetic hard axis of the Fe layer; however, spins away from the interface can rotate freely even in the low magnetic field.

# Introduction:

Exchange-coupled hard/soft ferromagnetic (FM) systems drew the attention of the scientific community due to their fascinating properties such as high energy density [1], exchange spring effect [1–3], exchange bias [4–7] etc., that have potential applications as permanent magnets as well as in magnetic recording media, these properties arise due to various interactions in the system, such as exchange interaction at the interface, dipolar interaction between adjacent domains, etc. [1–5,8]. A balance of these interactions is necessary to obtain desired properties for technological applications such as high thermal stability and low switching current of exchange coupled devices and high $BH_{max}$ for permanent magnets.

The strength of these interactions can be tuned by varying parameters, such as thickness. [6,9], magnetic anisotropy [10], composition, deposition condition, etc., of the individual layers. In particular, for hard/soft FM system with orthogonal magnetic anisotropy, i.e., the hard layer having perpendicular magnetic anisotropy and the soft layer having in-plane magnetic anisotropy, the dominant interaction between the layers can be tuned from exchange to dipolar by varying the thickness of the soft and hard layer [8,11–15]. With the variation of the strength of these interactions, the direction of orientation of spins in the layers also changes from in-plane to out-of-plane and vice versa. In the literature, various studies have discussed the variation of spin structure at the top interface when a soft (hard) layer is deposited on a hard (soft) layer. [6,11,12,16]. Laenens *et al.* [11] even traced the orientation of soft Fe layer moments with its increasing thickness at the top interface (away from the hard layer) using the grazing incident nuclear resonance scattering (GI-NRS) technique and a wedge Fe/FePt bilayer with the use of the $Fe^{57}$ marker layer. However, until today, the orientation of soft layer moments close to the interface is rarely discussed, though it is critical for fundamental understanding and from the technological application point of view. Using a theoretical model and one-dimensional micromagnetic simulation, respectively, Cain et al. [15] and Anh Nguyen et al. [12] reported a gradual rotation of spins along the thickness of the soft FM layer (hard FM layer) when deposited above the hard FM layer ( soft FM layer) for higher thicknesses. However, direct experimental evidence of orientation of the moments at the hard-soft interface is currently missing to the best of our knowledge.

In view of the above points, In the present work, we have studied spin orientation and variation of the strength of coupling with increasing thickness and depth of soft FM Fe layer in

[Co/Pt]$_{ML}$/Fe multilayer. [Co/Pt]$_{ML}$/Fe multilayer with OMA consists of a soft FM Fe layer with IMA and a hard Co/Pt multilayer with PMA. The thickness-dependent study uses the real-time in-situ magneto-optical Kerr effect (MOKE) technique. In-situ characterization not only provides genuine properties without surface contamination due to oxidation or capping layer but also increases the thickness of the same sample, keeping the microstructure and interface identical, thereby avoiding sample-to-sample variation. It also avoids any thickness gradient in individual layers, which might occur in wedge samples due to their considerable length.

Depth-dependent study along the soft FM layer is performed via the GI-NRS technique using two [Co/Pt]ML/Fe samples with Fe57 marker layers placed at the different interfaces. In one case, the Fe$^{57}$ marker layer was placed at the Co/Pt and Fe interface, whereas it was placed far away from that interface in another case. The measurements were done both in the as-deposited state and with varying external magnetic fields. GI-NRS measurement was performed because of its high sensitivity to the magnetization direction of the isotopic probe layer and depth selectivity [17]. From GI-NRS time spectra, the hyperfine field and relative orientation of the Fe magnetic moment ($\mu_{Fe}$) with respect to the incident wave vector are determined. In addition, the x-ray standing wave (XSW) technique under planar waveguide condition was utilized to enhance further the resonant counts from the Fe$^{57}$ isotope marker layer [18,19]. The study provides direct evidence of the spin structure at the marker layer position in contrast to other magnetic characterization techniques, which provide average information on the magnetic multilayer.

## Experimental:

Pt(~300 Å)/[Co(4.33 Å/Pt(30 Å)]$_{10}$ multilayers having PMA were deposited at room temperature (RT) using magnetron sputtering technique on Si (111) substrate covered with SiO$_2$ layer in a vacuum condition with a base pressure of $3\times10^{-7}$ mbar. The Pt buffer layer was used to enhance texturing along (111) crystallographic plane, which enhances PMA [20–23]. The Presence of PMA in the multilayer was confirmed using polar MOKE and magnetic force microscopy (MFM) techniques. Structural characterization was done using x-ray diffraction (XRD) and x-ray reflectivity (XRR). It may be noted that Pt/[Co/Pt]$_{10}$ multilayers will be denoted as [Co/Pt]$_{ML}$ for convenience throughout the paper. To study the coupling variation with soft FM layer thickness above this [Co/Pt]ML structure, the Fe layer was grown using the -beam evaporation technique in an ultra-high vacuum (UHV) chamber. The film was characterized during growth as a function of Fe layer thickness using in-situ MOKE. Reflection high energy electron diffraction (RHEED) measurements were also performed in situ before and after Fe layer deposition for structural characterization. This sample is denoted as [CoPt]$_{ML}$/Fe throughout the manuscript for convenience.

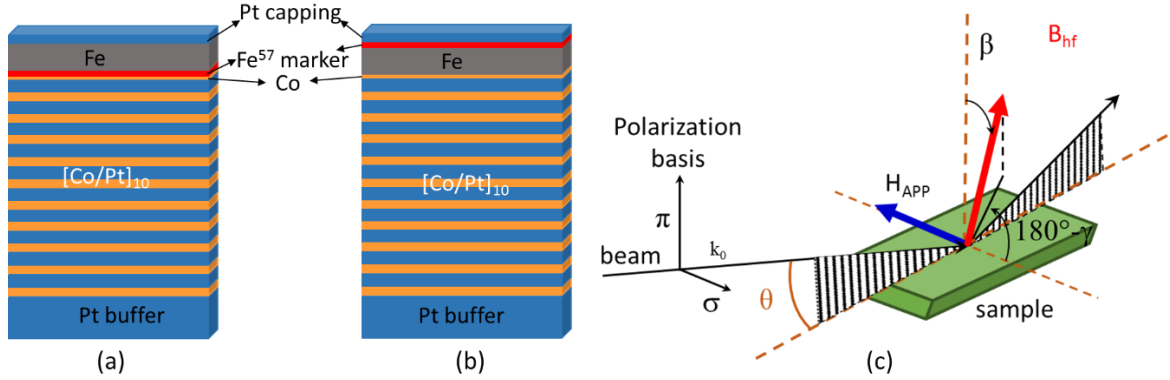

**Fig. 1:** Schematic of a multilayer structure of (a) Pt/[Co/Pt]$_{10}$/Co/Fe$^{57}$/Fe/Pt and (b) Pt/[Co/Pt]$_{10}$/Co/Fe/Fe$^{57}$/Pt showing the position of Fe$^{57}$ marker layer. (c) Schematic diagram showing the orientation of the sample with respect to the scattering plane of the incident beam and external applied magnetic field (H$_{APP}$). θ is the angle of incidence of the beam with respect to the sample surface,  is the angle of magnetization with respect to the sample normal, and γ is the direction of magnetization with respect to the scattering plane.

To further study the variation of strength of coupling and to probe the magnetic structure of the Fe layer when deposited on [Co/Pt]$_{ML}$ as a function of depth, two samples Pt/[Co/Pt]$_{10}$/Co/Fe$^{57}$/Fe/Pt and Pt/[Co/Pt]$_{10}$/Co/Fe/Fe$^{57}$/Pt with different position of Fe$^{57}$ marker layer (as shown in fig. 1(a) and 1(b)) were also prepared. The former multilayer has a Fe57 marker layer at [Co/Pt]ML - Fe interface (Fe bottom interface), whereas the latter one has a Fe57 marker layer at the Fe - Pt interface (Fe-top interface). Both samples were deposited at RT using the magnetron sputtering technique on Si (111) substrate. The thin Co (~4 Å± 0.2 Å) layer was deposited at the interface of [Co/Pt]$_{ML}$ and Fe to enhance PMA and avoid interdiffusion between Pt and Fe layers. The multilayer was capped with a high-density Pt layer to avoid surface oxidation and to generate XSW to enhance the NRS signal from the Fe$^{57}$ layer [24]. GI-NRS measurements were performed in an as-deposited state and with increasing external HAPP for both samples to study the coupling strength. The GI-NRS experiments used an x-ray synchrotron radiation source at the P01 Dynamics Beamline at PETRA III, DESY, Hamburg, Germany [25]. Experiments are conducted in time mode with a beam having 40 bunch with a bunch separation of 192 ns. The x-ray energy was tuned to 14.41 keV, the nuclear transition energy corresponding to the Fe$^{57}$ Mössbauer isotope with a natural lifetime of 141 ns. The grazing angle was decided based on the electronic (prompt with few nanoseconds) and nuclear (delayed 20–140 ns) reflectivity spectra to enhance the scattered radiation count.

The samples were mounted between the magnetic poles of an electromagnet such that external $H_{APP}$ is in the film plane and transverse to the direction of the incident beam ($k_o$) and the easy magnetic axis of the Fe layer (see Fig. 1c). The measured time spectra were fitted using the simulation and least-squares fitting procedure of the REFTIM program. [26].

## Result and discussion

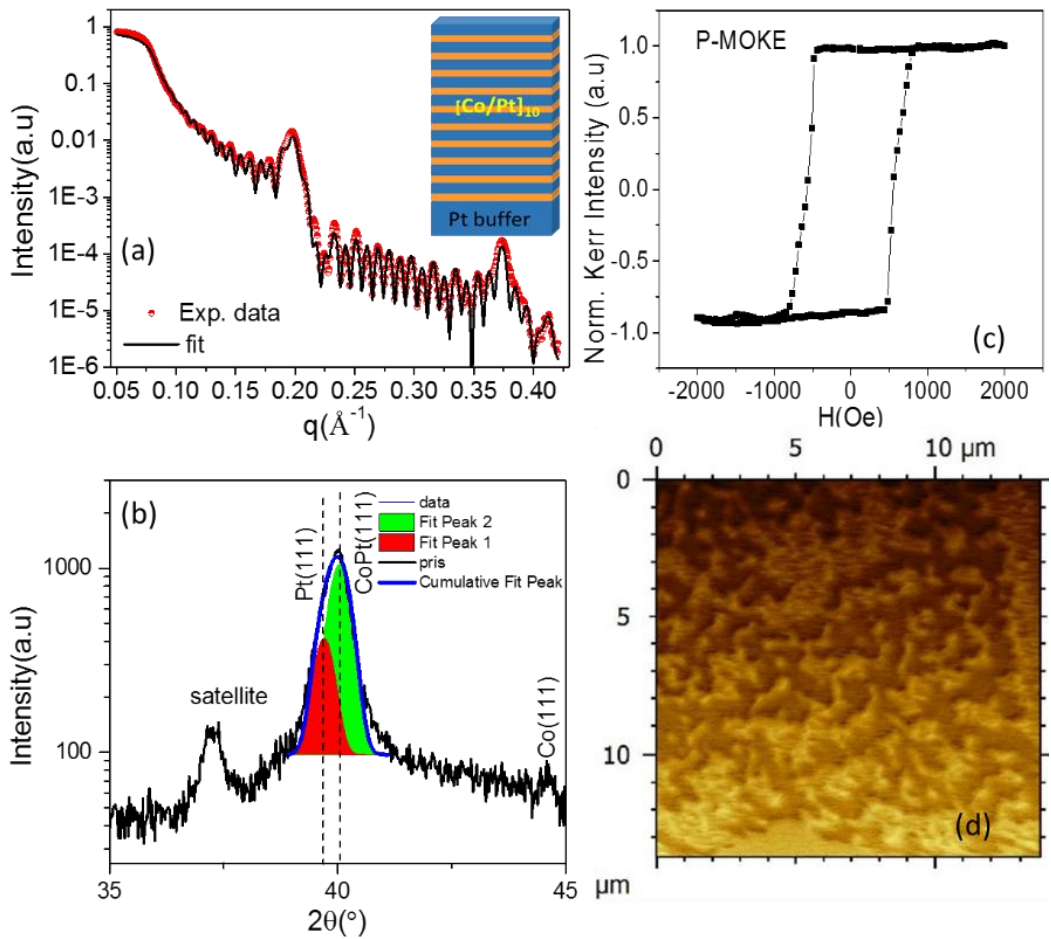

**Fig. 2:** (a) XRR pattern (red dots) of the $[Co/Pt]_{ML}$. The black line shows fit to the data obtained using Parratt's formalism Inset shows a schematic diagram of the $[Co/Pt]_{ML}$ and (b) XRD pattern of $[Co/Pt]_{ML}$ in θ-2θ geometry using Cu-K$_\alpha$. (c) MOKE hysteresis loop taken in polar geometry (polar -MOKE) and (d) MFM image of $[Co/Pt]_{ML}$.

Figure 2 shows (a) XRR pattern and corresponding (b) θ-2θ XRD pattern, (c) MOKE hysteresis loop obtained in polar geometry (P-MOKE), and (d) MFM image of [Co/Pt]$_{ML}$ multilayer. Inset of fig. 2a shows a corresponding schematic representation of the multilayer. XRR pattern was fitted using Parratt's formalism [27]. The fitted parameters are listed in table 1. Periodic small oscillations with high frequency correspond to the total thickness of the multilayer. Bragg peaks around q= 0.19 Å$^{-1}$ and 0.38 Å$^{-1}$ are observed due to the bilayer periodicity of the Co/Pt multilayer [28]. The presence of first and second-order Bragg peaks confirms the formation of a multilayer with a smooth interface.

**Table 1:** XRR fitting parameters of [Co/Pt]$_{ML}$ using Parratt's formalism [27]. The error in thickness and roughness is ± 0.5 Å.

| Element | Thickness, d (Å) | Roughness, σ (Å) | Scattering length density (SLD), ρ (Å$^{-2}$) |
|---|---|---|---|
| Co | 4.3 | 6.8 | 4.622E-05 |
| Pt | 30.0 | 5.8 | 1.397E-04 |
| Pt buffer | 305.0 | 4.5 | 1.299E-04 |

A broad peak in the XRD pattern around θ = 40° was observed, which can be de-convoluted into two peaks located at θ = 39.7° and 40.04° respectively. A peak at θ = 40.04° corresponds to the face-centered cubic (fcc) (111) plane of CoPt (111), whereas a peak at θ = 39.7° corresponds to fcc Pt (111) from Pt buffer. Strong texturing of [Co/Pt]$_{ML}$ along (111) planes normal to the surface is due to the Pt buffer layer [20–23]. As per literature, the peak corresponding to [Co/Pt]$_{ML}$ lies between fcc Pt(111) and fcc Co(111) peak, and its position depends on the thickness of Pt and Co layer and strain in the lattice of both Pt and Co layers [20–23]. Since the thickness of Pt is large as compared to the Co layer in the Co/Pt multilayer, the CoPt(111) peak in the present case lies close to the Pt(111) peak position (θ = 39.7°) [29–31]. This results in an overlap of the CoPt (111) peak with the (111) peak of the Pt buffer. A peak at 37.2° is a satellite peak corresponding to the superlattice bilayer period in [Co/Pt]$_n$ multilayer [32]. Satellite peak arises due to structural coherency between the Co and Pt layers. The presence of the fcc (111) plane in the XRD pattern indicates PMA in the [Co/Pt]$_{ML}$ [21,23,33].

Almost square hysteresis loop with very high remanence (Mr=0.94) and low coercivity ($H_C$ ~560 Oe) obtained in polar geometry (P-MOKE) clearly shows that the $[Co/Pt]_{ML}$ multilayer exhibits strong PMA. The origin of PMA in this multilayer is known to originate from the spin-orbit coupling between Co and Pt atoms at the interface [22,33,34]. In the MFM image, the presence of large labyrinth domains in the demagnetized state further confirms the Presence of PMA in $[Co/Pt]_{ML}$ [35–38].

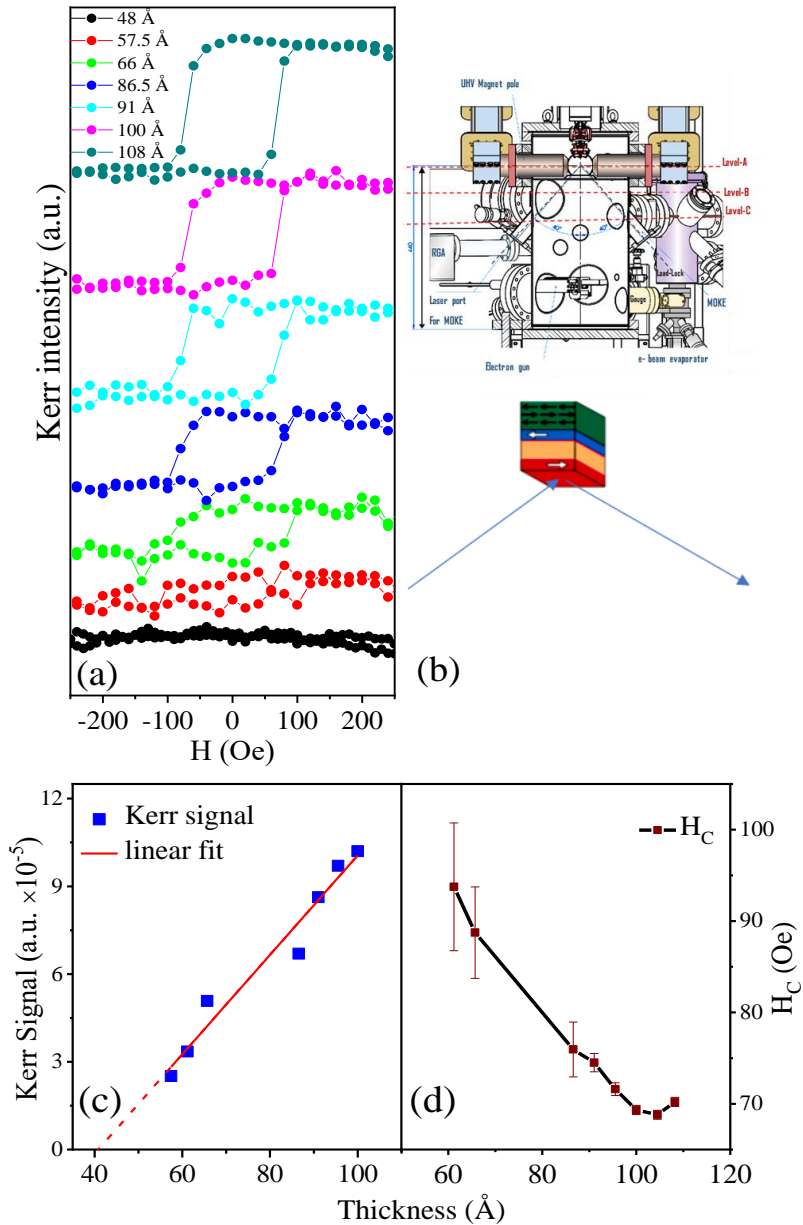

**Fig. 3:** (a) MOKE hysteresis loop as a function of Fe layer deposition on $[Co/Pt]_{ML}$ layer. (b) displays the schematic of the *in-situ* MOKE setup. Variation of (c) Kerr signal and (d) $H_C$ as a function of Fe layer thickness.

Figure 3(a) shows the hysteresis loop of the Fe layer deposited on [Co/Pt]$_{ML}$ as a function of increasing thickness, as obtained using the in-situ MOKE technique in longitudinal geometry (L-MOKE) during growth. A clear change in the magnetic signal of the hysteresis loops was observed with increasing thickness. Upto 48 Å hysteresis loop was absent, whereas on increasing thickness to ~57 Å, a small hysteresis loop was obtained, further developing into a square hysteresis loop with an increase in Fe layer thickness. It may be noted that the magnetic field used in MOKE measurements was kept at about ±250 Oe. This field was insufficient to influence the magnetization of underneath [Co/Pt]$_{ML}$ template layer, which is magnetically hard and has strong PMA. The Fe layer is magnetically soft and exhibits in-plane magnetization reversal during MOKE measurement. Therefore, hysteresis loops obtained in the present thickness-dependent measurements only correspond to the Fe layer. Fig. 3(c) and 3(d) give variations in the Kerr signal (height of the hysteresis loop) and $H_C$ with increasing Fe thickness. Kerr signal increases almost linearly with increasing thickness, whereas $H_C$, first decreases and then saturates at around 85 Å thickness. A linear fit to the variation of the Kerr signal as a function of thickness gives a magnetic dead layer or inactive layer of ~42 Å [39,40]. The presence of an inactive layer at the interface can be explained in terms of competition between exchange energy and magnetostatic energy [8,11–14]. It is to be further noted that in contrast to collinear hard/soft FM system with in-plane magnetic anisotropy [4,5,41] where deposition of soft FM layer on hard FM layer results in shifted hysteresis loop, in the present case, all hysteresis loops are centered along the field axis, i.e., EB is absent. The absence of EB in Fe film can be explained by the absence of any preferential pinning direction due to hard [Co/Pt]$_{ML}$ during the magnetization reversal of the Fe layer.

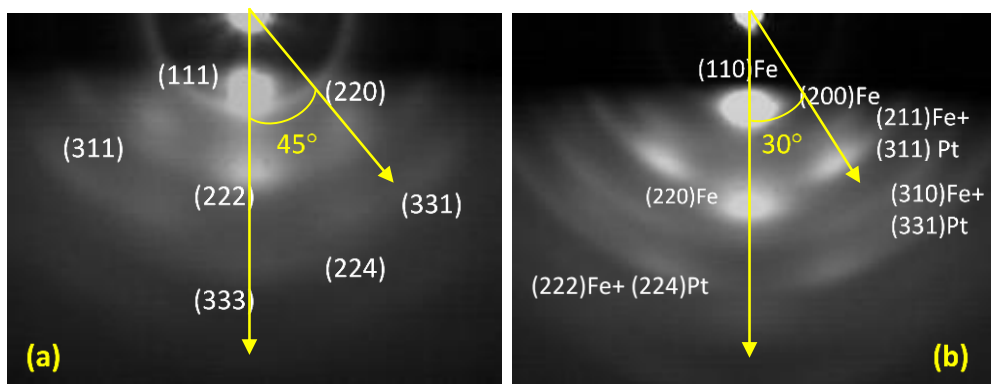

**Fig. 4:** RHEED pattern of [Co/Pt]$_{ML}$/Fe (a) before and (b) after Fe layer deposition.

Figures 4(a) and 4(b) show RHEED patterns of Co/Pt multilayer before and after deposition of the Fe layer, respectively. Concentric rings (Debye rings) show that the film is polycrystalline. Planes corresponding to the rings are designated in the RHEED pattern. It was observed that new planes corresponding to the Fe layer appear in Fig. 4(b). Discontinuity in the arc of the ring in both images, before and after the Fe layer deposition, shows that the films exhibit texturing. However, as compared to the Co/Pt multilayer, the Fe layer has weak texturing. Strong texturing of fcc (111) planes of Co/Pt multilayer along 90° direction, i.e., normal to the substrate surface, was observed before Fe deposition, whereas fcc (220) planes are textured along 45° with respect to the surface normal. On the deposition of the Fe layer, (110) planes of Fe show strong texturing along the surface normal, whereas the (211) plane is textured along 30° with respect to the surface normal. Strong texturing of the (111) plane in the Co/Pt multilayer RHEED pattern in the out-of-plane direction further confirms the strong PMA behavior of the Co/Pt multilayer [23,33].

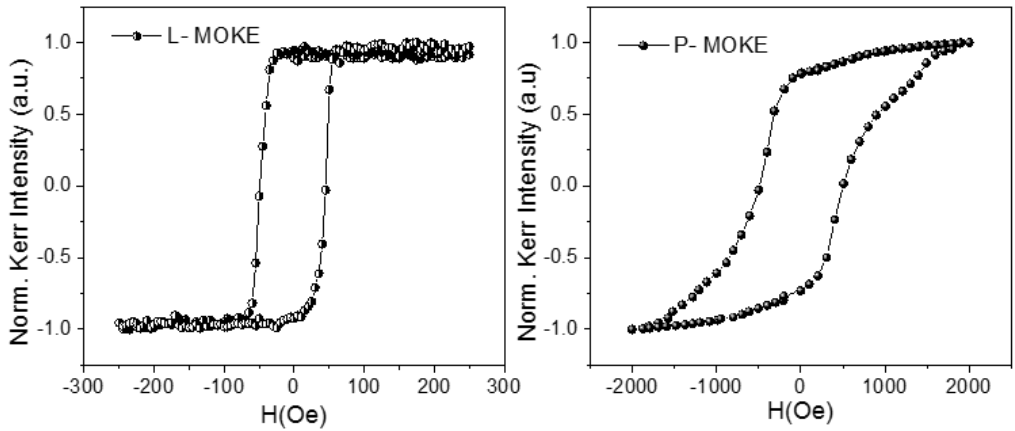

**Fig. 5: (a)** L-MOKE and (b) P-MOKE hysteresis loop of Pt/[Co/Pt]$_{10}$/Co/Fe$^{57}$/Fe/Pt.

Figure 5 shows the hysteresis loop of Pt/[Co/Pt]$_{10}$/Co/Fe$^{57}$/Fe/Pt in longitudinal (L) and polar (P) MOKE geometry. It was observed that the L-MOKE hysteresis loop exhibits almost square loop, whereas the P-MOKE loop is a combination of two loops, where the square loop corresponds to the Co/Pt multilayer and the slanted loop corresponds to the Fe layer with an easy axis along the in-plane direction. In the case of the L-MOKE, a single loop was observed due to large PMA in [Co/Pt]$_{ML}$. Hence, ±the 250 Oe field used in L-MOKE is sufficient for the magnetization reversible for the soft magnetic Fe layer but much less than the field required to [Co/Pt]$_{ML}$. On the other hand, in the case of P-MOKE, the magnetic easy axis of [Co/Pt]$_{ML}$ lies along the direction of the applied field, and therefore, magnetic reversal takes place in a relatively much smaller field. Still, the Fe layer needs a larger field, about 2000 Oe, for magnetization reversal in this geometry. Because of this fact, both the Fe layer and [Co/Pt]$_{ML}$ multilayer participate in the magnetic reversal process with field strength ±2000 Oe and the same is also visible in terms of combined two loops in P-MOKE (fig. 5(b)).

GI-NRS measurement was performed on [Co/Pt]$_{ML}$/Fe multilayer with very thin Fe$^{57}$ marker layers to study the strength of coupling at the top and bottom interface of the Fe layer due to its sensitivity to the Fe$^{57}$ isotope. However, the thickness of the marker layer is very small. Hence, to enhance the delayed counts from the Fe$^{57}$ marker layer, the x-ray radiation has been coupled into a XSW mode. In the present case, XSW is generated through grazing incidence total external reflection method where a low-density guiding layer Co/Fe$^{57}$/Fe is sandwiched in between two high-density Pt layers - **Pt**/Co/Fe$^{57}$/Fe/**Pt** acting as walls of the planar waveguide. Resonance modes are excited in between the critical angle of the substrate and the high-density layer due to multiple interference resulting from total external reflection [42–45].

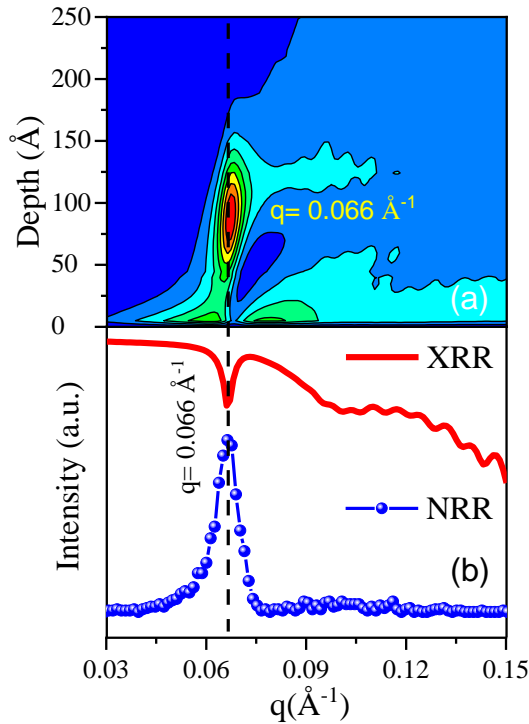

**Fig. 6:** (a) Contour plot of x-ray field intensity inside Pt/[Co/Pt]10/Co/Fe$^{57}$/Fe/Pt multilayer as a function of depth and scattering vector q. At q=0.066 Å$^{-1}$, the XSW antinode (TE0 mode) crosses the Fe$^{57}$ marker layer. (b) displays the multilayer's experimental non-resonant (XRR) and resonant reflectivity (NRR), showing minima and maxima, respectively, at q=0.066 Å$^{-1}$.

Based on the final structure, such as thickness, roughness, etc., as obtained from the x-ray reflectivity (fig. 6(b)), x-ray field intensity inside the Pt/[Co/Pt]$_{10}$/Co/Fe$^{57}$/Fe/Pt multilayer structure was calculated as a function of scattering vector q using the Parratt's formalism [27] and shown in fig. 6(a). Clear confinement of the x-ray field intensity (TE0 mode) was observed at around the angle of incidence θ=0.259° (q=0.066Å$^{-1}$). From the calculated electric field intensity at q= 0.066 Å$^{-1}$ to the position of the Fe$^{57}$ marker layer, it is found that the standing wave antinode (TE0 mode) crosses the Fe$^{57}$ layer. It confirms that nuclear resonance counts in NRS measurement will increase if the measurements are done by keeping the incident angle fixed at q= 0.066 Å$^{-1}$. Nuclear resonance counts are collected as a function of q and presented in Fig. 6(b) to confirm this. A clear 40-fold increase in the intensity has been observed at around q=0.066 Å$^{-1}$. Because of the above observations, all NRS measurements (time spectra) were recorded at this q value with maximum resonance count.

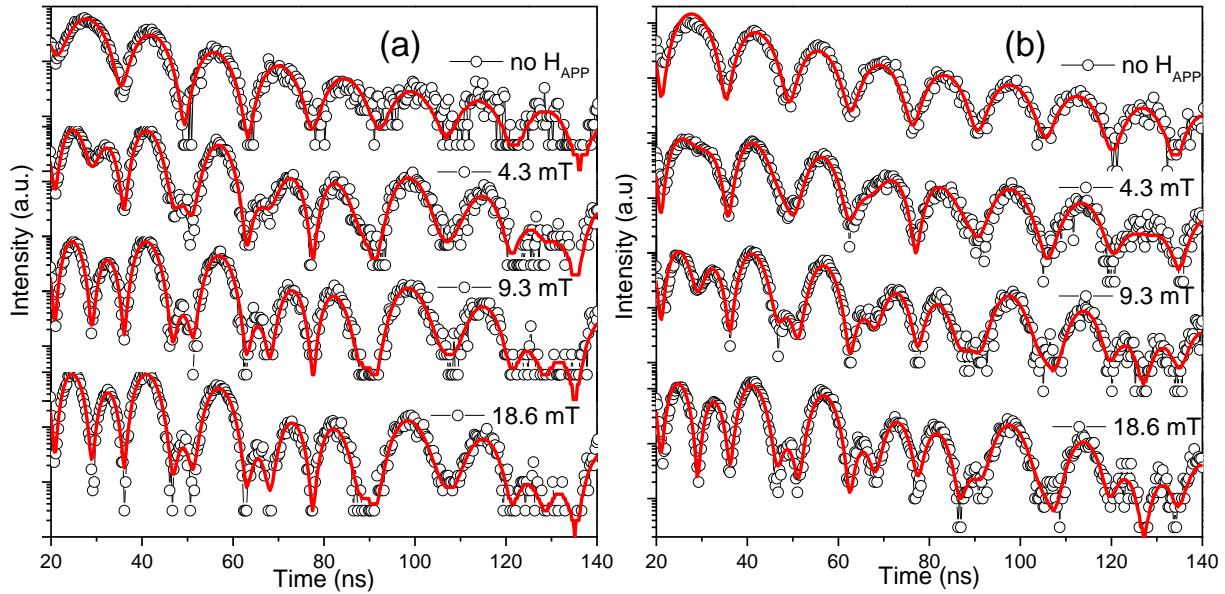

**Fig. 7:** GI-NRS time spectra of (a) Pt/[Co/Pt]$_{10}$/Co/Fe/Fe$^{57}$/Pt and (b) Pt/[Co/Pt]$_{10}$/Co/Fe$^{57}$/Fe/Pt with varying strength of external H$_{APP}$. Dots represent the experimental data, and lines show the best fit for the data obtained using REFTIM software.

Figure 7 shows GI-NRS time spectra of (a) [Pt/[Co/Pt]$_{10}$/Co/Fe/Fe$^{57}$/Pt and (b) Pt/[Co/Pt]$_{10}$/Co/Fe$^{57}$/Fe/Pt in the as-deposited state (no H$_{APP}$) as well as with varying strength of external H$_{APP}$. As the NRS technique can detect the Fe57 isotope layer only, all the time spectra correspond to the Fe$^{57}$ layer at the interface. The direction of H$_{APP}$ was kept normal to the scattering plane of the incident beam and the easy axis of the Fe layer. The spectra were fitted using REFTIM software [26] and the Fitting parameters obtained are tabulated in Table 2. Structural parameters for the initial fitting were obtained from XRR analysis. It was observed that for the best fit to the time spectra, three hyperfine field (Bhf) components were to be considered for the $Fe^{57}$ layer.

**Table 2**. Fitting parameters of GI-NRS using REFTIM [26].

| $H_{APP}$ (mT) | [Pt/[Co/Pt]$_{10}$/Co/Fe/Fe$^{57}$/Pt] | | | Pt/[Co/Pt]$_{10}$/Co/Fe$^{57}$/Fe/Pt | | |
|---|---|---|---|---|---|---|
| | Bhf(T) | β(°) | γ(°) | Bhf(T) | β(°) | γ(°) |
| **0** | 33.02 | 90.6 | 90.0 | 32.89 | 2.0 | 2.0 |
| | 31.59 | 90.0 | 90.0 | 32.61 | 0.0 | 2.0 |
| | 32.16 | 83.2 | 90.0 | 28.75 | 0.0 | 2.0 |
| **4.3** | 33.02 | 90.0 | 16.5 | 32.71 | 45.2 | 0.0 |
| | 31.59 | 90.0 | 31.2 | 32.49 | 51.1 | 0.0 |
| | 32.16 | 90.0 | 15.1 | 28.97 | 43.5 | 0.0 |
| **9.3** | 33.02 | 90.0 | 3.5 | 32.47 | 72.2 | 0.0 |
| | 31.59 | 90.0 | 15.2 | 32.66 | 69.2 | 0.0 |
| | 32.16 | 90.0 | 0.1 | 28.49 | 66.2 | 0.0 |
| **18.6** | 33.02 | 90.0 | 3.5 | 32.46 | 90.0 | 0.0 |
| | 31.59 | 90.0 | 1.2 | 32.66 | 90.0 | 0.0 |
| | 32.16 | 90.0 | 0.1 | 28.49 | 90.0 | 0.0 |

In the absence of an external field, both time spectra exhibit quantum beats with a single oscillation frequency but different amplitudes [46]. After fitting, it was found that spins of the $Fe^{57}$ marker layer in Pt/[Co/Pt]$_{10}$/Co/Fe$^{57}$/Fe/Pt are oriented normally to the surface, whereas in the case of Pt/[Co/Pt]$_{10}$/Co/Fe/Fe$^{57}$/Pt sample, spins are aligned in the film plane. This shows that the orientation of Fe layer moments varies along its depth. This behavior is due to variations in the coupling strength along the depth of the Fe layer.

It is further observed that when $H_{APP}$ ~ 4.3 mT was applied, the time spectra for both the films showed splitting of oscillations. However, splitting was more prominent for

[Pt/[Co/Pt]$_{ML}$/Co/Fe/Fe$^{57}$/Pt than Pt/[Co/Pt]$_{ML}$/Co/Fe$^{57}$/Fe/Pt. On further increasing the strength of H$_{APP}$, complete splitting of oscillation in the time spectra occurs. For [Pt/[Co/Pt]$_{ML}$/Co/Fe/Fe$^{57}$/Pt, complete splitting is observed at low field (9.3 mT), whereas for Pt/[Co/Pt]$_{ML}$/Co/Fe$^{57}$/Fe/Pt complete splitting was observed at higher field (18.6 mT). It shows that a high field is required to rotate spins of the Fe layer close to the interface in the direction of the H$_{APP}$, whereas about 9.3 mT field is sufficient to align the moments away from the interface.

**Discussion:**

At very low thickness, exchange energy will dominate [8,11–14]plane direction. Consequently, the Fe layer does not participate in magnetization reversal within the in-plane H$_{APP}$ range and behaves as a magnetically inactive layer. Thus, no hysteresis was observed up to 48 Å. With increasing thickness, this exchange energy becomes weak. Hence, under the effect of magnetostatic energy, moments of the Fe layer away from the interface start to rotate in the in-plane direction to form flux closure domains between the perpendicular domains of the Co/Pt multilayer [8,11–14]. This results in net in-plane magnetization in the Fe layer. The initial canting of loops was due to the moments of the Fe layer that had PMA but was forced to rotate in an in-plane direction under H$_{APP}$. The appearance of a square loop with a further increase in thickness was due to the in-plane moments of the Fe layer, which became independent of the Co/Pt multilayer due to weak coupling with Fe atoms away from the interface. The decrease of H$_C$ with increasing thickness of the Fe layer is again due to the weakening of coupling. This behavior is in contrast to the variation in H$_C$ of soft FM when deposited on the nonmagnetic template, where H$_C$ initially increases and reaches the maximum at about ~100 Å thickness. Further increase in thickness results in a decrease in H$_C$, which is attributed to the reduction of domain walls pinning at the surface [39]

The absence of EB in Fe film can be explained by the absence of any preferential pinning direction due to hard [Co/Pt]$_{ML}$ during the magnetization reversal of the Fe layer. Asymmetric domain pining by hard FM multilayer is the required condition to observe EB in such magnetic structures [47,48]

This behavior is due to variation in the coupling strength along the depth of the Fe layer, which is strong at the interface of perpendicularly magnetized [Co/Pt]ML and Fe layers, resulting in perpendicular alignment of Fe layer moments. The coupling decreases as we move away from the interface region. In the present case, for Pt/[CoPt]$_{ML}$/Co/Fe$^{57}$/Fe/Pt multilayer, the Fe$^{57}$ marker layer is close to the interface; hence, it is strongly coupled to the [Co/Pt]$_{ML}$, and its moments are aligned normal to the film plane. However, in the case of [Pt/[CoPt]$_{ML}$/Co/Fe/Fe$^{57}$/Pt multilayer, the Fe$^{57}$ marker layer is away from the interface; hence, the exchange coupling between [Co/Pt]$_{ML}$ and Fe$^{57}$ moments is weak. This leads to the in-plane alignment of Fe moments due to the magnetostatic stray field of [Co/Pt]$_{ML}$. This is in accordance with the study of Anh Nguyen *et al*. [12] [**Error! Bookmark not defined.**], where they observed that the net magnetization of the soft FM NiFe layer rotates from out of plane at the interface to towards in-plane direction away from the interface when deposited on hard FM [Co/Pd]$_5$ multilayer. They found that the thickness of the NiFe layer can tune the tilt angle at the top interface.

This further evidences that moments at the interface are strongly coupled by [Co/Pt]$_{ML}$ compared to moments away from the interface. Hence, the strength of coupling varies within the soft FM layer.

## Conclusion:

In the present work, spin orientation and variation of the strength of coupling in [Co/Pt]$_{ML}$/Fe multilayer as a function of thickness and depth of Fe layer has been investigated using in-situ magneto-optical Kerr effect (MOKE) and isotope sensitive depth-resolved nuclear resonance scattering (NRS) technique. In-situ MOKE measurement as a function of Fe layer deposition reveals that with the increase in thickness of the Fe layer, moments of Fe layer reorientation from out-of-plane to in-plane direction as a result of the decrease in the coupling between [Co/Pt]$_{ML}$ and Fe layer. Spin orientation and variation of the strength of coupling as a function of depth for 75Å Fe layer deposited on [Co/Pt]$_{ML}$ is studied via NRS technique using two [Co/Pt]$_{ML}$/Fe multilayers having the same thickness but different position of Fe$^{57}$ marker layer. Films were studied in their deposited state and with varying external magnetic fields. It is observed that the coupling varies along the depth of the Fe layer, strong at [Co/Pt]$_{ML}$ and Fe interface with spins of the Fe layer aligned normally to the film plane, whereas moments away from the interface are weakly coupled and aligned in-plane along the magnetic easy axis. Due to this gradient in coupling strength along depth, relatively higher $H_{APP}$ is required to reorient Fe spins near the Co/Pt layer, whereas spins away from the interface rotate freely with relatively weak $H_{APP}$.

**Acknowledgement:**

We acknowledge Dr. V. Raghavendra Reddy and Er. Anil Kumar, UGC-DAE CSR, Indore, India, for polar MOKE measurement. We acknowledge Deutsches Elektronen-Synchrotron (DESY) (Hamburg, Germany), a member of the Helmholtz Association HGF, for providing experimental facilities. Parts of this research were carried out at PETRA III, and we would like to thank Kai Schlage for assistance in using beamline P01, PETRA III. I would also like to thank the Department of Science and Technology (DST), Government of India, for providing financial assistance to perform experiments at DESY, Hamburg, within the framework of the India@DESY collaboration (Proposal No. I-20160350). We acknowledge Er. Layant Behera and Prabhat Kumar, UGC-DAE CSR, Indore, India, for film deposition using sputtering technique and XRD measurements. We acknowledge Dr. Pooja Gupta and Dr. P. Nageshwararao, RRCAT, Indore, India, for MFM measurement.